\documentclass[pdflatex,sn-mathphys-num,referee]{sn-jnl}

\usepackage{physics,amsmath,soul,color, amssymb, cancel, amsthm, bbm}
\usepackage{wasysym, enumitem, graphicx, wrapfig, multirow, makecell, float, hyperref, hhline, booktabs}
\usepackage{subcaption, tikz, mwe}
\usepackage[normalem]{ulem}

\DeclareMathAlphabet{\mathpzc}{OT1}{pzc}{m}{it}
\usetikzlibrary{shapes,arrows,positioning}

\DeclareMathSizes{10}{9.5}{6}{6}
\makeatletter
\g@addto@macro \normalsize {%
 \setlength\abovedisplayskip{5pt plus 2pt minus 2pt}%
 \setlength\belowdisplayskip{5pt plus 2pt minus 2pt}%
}
\makeatother

\newcolumntype{C}[1]{>{\centering\arraybackslash}m{#1}}
\newcolumntype{P}[1]{>{\centering\arraybackslash}p{#1}}
\newcolumntype{M}[1]{>{\centering\arraybackslash}m{#1}}

\newcommand{\CalI}{\mathcal{I}}
\newcommand{\CalJ}{\mathcal{J}}

\newcommand{\bfth}{\pmb{\theta}}

\newcommand{\bfw}{\mathbf{w}}
\newcommand{\bfx}{\mathbf{x}}

\begin{document}

\title{Inverse design of bespoke interatomic potentials via active learning by information-matching}

\author[1]{\fnm{Yonatan} \sur{Kurniawan}}\email{kurniawanyo@outlook.com}
\author[2]{\fnm{Logan D.} \sur{Williams}}\email{williams332@llnl.gov}
\author[2]{\fnm{Amit} \sur{Samanta}}\email{samanta1@llnl.gov}
\author[3]{\fnm{Ilia} \sur{Nikiforov}}\email{nikif002@umn.edu}
\author[4]{\fnm{Daniel} \sur{Schwalbe-Koda}}\email{dskoda@ucla.edu}
\author[5]{\fnm{Mark K.} \sur{Transtrum}}\email{mark@cross-stream.ai}
\author[3]{\fnm{Ellad B.} \sur{Tadmor}}\email{tadmor@umn.edu}
\author[2]{\fnm{Vincenzo} \sur{Lordi}}\email{lordi2@llnl.gov}
\author*[2]{\fnm{Vasily V.} \sur{Bulatov}}\email{bulatov1@llnl.gov}

\affil[1]{\orgdiv{Department of Physics and Astronomy}, \orgname{Brigham Young University}, \orgaddress{\city{Provo}, \state{UT}, \country{USA}}}
\affil[2]{\orgname{Lawrence Livermore National Laboratory}, \orgaddress{\city{Livermore}, \state{CA}, \country{USA}}}
\affil[3]{\orgdiv{Department of Aerospace Engineering and Mechanics}, \orgname{University of Minnesota}, \orgaddress{\city{Minneapolis}, \state{MN}, \country{USA}}}
\affil[4]{\orgdiv{Department of Materials Science and Engineering}, \orgname{University of California}, \orgaddress{\city{Los Angeles}, \state{CA}, \country{USA}}}
\affil[5]{\orgname{Cross Stream Consulting}, \orgaddress{\city{Springville}, \state{UT}, \country{USA}}}

\abstract{
    Interatomic potentials (IPs) enable large-scale atomistic simulations beyond the reach of first-principles methods, but their predictive reliability depends critically on the selection of training data, quantified uncertainty, and model expressiveness.
    Active learning (AL) provides a principled framework for constructing efficient and accurate IPs, yet most strategies reduce parameter uncertainty without explicitly accounting for the specific material properties being predicted.
    The information-matching (IM) approach addresses this limitation by requiring that the selected training data provide at least as much parameter space information as needed to achieve prescribed uncertainty targets for selected quantities of interest (QoIs).
    Here, we apply IM to develop bespoke IPs specifically tailored for predicting plastic strength in metals.
    Due to the high computational cost of simulating plastic strength, we employ an indirect IM strategy that targets inexpensive intermediate QoIs that correlate with strength.
    The IM method enables precise parameter constraints with minimal training data, yielding precise predictions for both the intermediate QoIs and plastic strength.
    Yet, model error remains a key limitation, and a post hoc uncertainty inflation correction provides a viable means to mitigate this limitation.
    These findings illustrate both the promise and limits of uncertainty-aware AL for predicting complex material properties.
}


\maketitle

\section{Introduction}
\label{sec:introduction}

Atomistic simulations provide a powerful framework for probing the fundamental mechanisms that govern material behavior.
These simulations offer direct insight into processes such as defect formation, plastic deformation, and phase transitions, which are difficult to access experimentally at the relevant length and time scales.
A central component of such simulations is the interatomic potential (IP), which encodes the effective interactions between atoms and determines the accuracy of the predicted material properties.
While first-principles approaches, such as density functional theory (DFT), offer high accuracy, their computational cost is prohibitive for the large systems and long timescales required to study the mechanical response of materials.
In contrast, IPs provide a computationally efficient alternative that captures the essential physics of bonding and interactions, enabling simulations of millions of atoms and bridging the gap toward experimentally relevant conditions.

Generally, the development of an IP model requires prioritizing among the trilemma of mutually conflicting requirements:
(1) model transferability across different material systems and properties of interest,
(2) model accuracy, and
(3) model computational efficiency.
One approach that has gained popularity in recent years is the construction of \emph{universal} or \emph{foundation} IPs designed to work across a wide range of materials and conditions \cite{chen_universal_2022,batzner_e3-equivariant_2022,deng_chgnet_2023,batatia_foundation_2024}.
An alternative strategy, which we pursue here, is to generate \emph{bespoke} or \emph{\`{a} la carte} IPs, that sacrifice broad transferability in favor of greater accuracy and efficiency for a specific targeted application \cite{alavi_ani-1ccx-gelu_2025,radova_fine-tuning_2025,liu_study_2025}.

Regardless of its intended uses, the predictive accuracy of an IP is strongly dependent on the quality and diversity of the training data.
A straightforward approach is to generate a large ground truth (GT) training dataset, but this approach can become computationally expensive due to the typically high acquisition cost of GT data.
Additionally, high degrees of redundancy in arbitrary large datasets often lead to training bias and less accurate IPs.
Active learning (AL) addresses this challenge by iteratively identifying and selecting the most informative data points for training \cite{leardi_experimental_2009,alizadeh_survey_2021}.
In the context of atomistic simulations, AL is often coupled with on-the-fly learning, where GT data are selected and generated dynamically during simulations \cite{podryabinkin_active_2017,jinnouchi_--fly_2020}.
This strategy ensures that the potential is continually refined in regions of configuration space that are most relevant to the target application.
As a result, the number of GT calculations can be reduced substantially, leading to significant savings in computational cost without substantially sacrificing model accuracy.

In addition to achieving high prediction accuracy, it is important to assess the reliability and precision of predictions based on an IP model.
As a surrogate model, the reliability of an IP depends on the simulation conditions under which it is applied.
To provide meaningful information alongside experimental data, predictions from an IP should be accompanied not only by an estimate of its prediction error, but also by a quantified prediction uncertainty.
While error measures accuracy with respect to GT, uncertainty characterizes the model’s confidence in its predictions.
In many practical applications, including the one considered in this work, direct evaluation of prediction error by validation against GT is infeasible because of the prohibitive cost of GT calculations.
In such cases, prediction uncertainty has been suggested to serve as a practical proxy for prediction error \cite{frederiksen_bayesian_2004,kurniawan_comparative_2025}.
In this context, having a large but known prediction uncertainty may be more useful than having unrecognized or unknown uncertainty, as it enables informed assessment of model reliability and supports uncertainty-aware decision making.

Recently, we introduced a novel AL strategy, the \emph{information-matching} (IM) method, that incorporates prediction uncertainty directly into the data selection process \cite{kurniawan_information-matching_2026}.
Like some other AL approaches \cite{jacroux_-optimality_1989,morgan_e-optimality_2011,jones_-optimal_2021}, IM relies on the Fisher information matrix (FIM) to quantify the information content of IP parameters.
However, rather than aiming to minimize parameter uncertainty indiscriminately and irrespective of the IP's intended use, IM selects training data by requiring that the information supplied by the data is sufficient to achieve prescribed uncertainty targets for specific quantities of interest (QoIs).
The IM method identifies the level of IP parameter precision required to predict the target QoI with the desired precision and then selects a minimal set of training data sufficient to meet this requirement.
Thus, IM avoids unnecessary overspecification of the IP parameters while ensuring that model reliability is aligned with the prediction task.

In previous work \cite{kurniawan_information-matching_2026}, we demonstrated the versatility of the IM method across distinct experimental design problems, including sensor placement in power networks and underwater acoustics, and construction of an IP for targeted material QoIs.
In the latter case, specifically, the QoIs were relatively simple material properties that could be computed from and validated against ground truth DFT, thus providing a useful test bed for the method.
In this work, we push the IM method into a far more demanding setting, aiming to develop an IP capable of predicting the plastic strength of a metal.
Unlike previously considered simple material properties, plastic strength emerges from the collective dynamics of extended defects such as dislocations and deformation twins.
Capturing these microstructural mechanisms requires molecular dynamics (MD) simulations on extreme scales---on the order of $10^8$ atoms and $10^6$ integration time steps---far out of reach for direct validation against DFT.

Facing the ``direct validation being impossible'' conundrum, we investigate whether the prediction uncertainty of IP models developed using the IM method can serve as a meaningful proxy for the expected error in strength predictions. 
We take an indirect approach in which informative data for IP fitting are selected in an IM loop that, rather than targeting the ultimate QoI (strength) itself, targets several QoIs that are less costly to compute, referred to as indicator properties, that can be validated against GT calculations and are known to be covariant with strength.
In developing this indirect version of the IM method, the key question is whether constraining IP model parameters through these indicator properties delivers sufficiently confident predictions not only for the indicator properties themselves, but also for the ultimate target QoI, strength.

The remainder of the paper is organized as follows.
Section~\ref{sec:methods} describes the functional form of IP employed in this study and the IM algorithm used to select informative GT data and to train the IP.
Section~\ref{sec:results} presents results from several case studies, each contributing a step toward the overarching goal of constructing an IP model capable of predicting plastic strength.
Finally, Section~\ref{sec:discussion} discusses limitations of our approach and proposes possible corrections and extensions.
A summary of this work is then given in Sec.~\ref{sec:conclusion}.


\section{Methods}
\label{sec:methods}

Our goal in this work is to develop bespoke IPs optimized for predicting plastic strength of the metal tantalum (Ta) using an AL framework based on the IM method.
The IM workflow for IP optimization starts by defining three key ingredients:
(1) an IP functional form,
(2) a pool of candidate data points to be subsampled for IP training, and
(3) one or more target QoIs along with their desired uncertainty bounds.
In what follows, we first present and motivate the IP functional form adopted in this work, then define the loss function employed in IP training and describe two datasets that serve as candidate pools for AL.
Next, we introduce the IM method in detail and explain how it is embedded within the AL loop.
Finally, we describe our target QoIs and define their desired prediction uncertainties.
Using an iterative AL workflow with IM, we determine a minimal set of training data required to sufficiently constrain the IP parameters and to ensure that prediction uncertainties of the target QoIs remain within the prescribed bounds.

\subsection{EAM potential}
\label{sec:eam-potential}

For this study, we employ an embedded-atom method (EAM) potential developed in [\citenum{zhou_misfit-energy-increasing_2004}] as the functional form of the trial interatomic potential (IP).
This choice is motivated by three factors:
(i) the demonstrated success of EAM potentials in molecular dynamics (MD) simulations of plastic deformation in metals \cite{zepeda-ruiz_probing_2017,zepeda2021atomistic},
(ii) the computational efficiency of EAM potentials, and
(iii) the relatively small number of fitting parameters---specifically, 20 parameters in the EAM implementation proposed by Zhou et al. \cite{zhou_misfit-energy-increasing_2004}---which reduces the computational burden of IM-based IP development for metal strength simulations.

In the EAM formalism, the total energy of an atomic configuration with $N$ atoms is given by
\begin{equation}
    \label{eq:energy}
    E = \sum_{\substack{i, j=1 \\ j > i}}^N \phi (r_{ij}) + \sum_{i=1}^N F (\rho_i),
\end{equation}
where $\phi (r_{ij})$ is the energy of interaction between a pair of atoms $i$ and $j$ separated by distance $r_{ij}$, and $F (\rho_i)$ is the energy produced by embedding atom $i$ into the local electron density $\rho_i$ supplied by all other atoms.
The local electron density at the position of atom $i$ is calculated as
\begin{equation}
    \label{eq:electron_density}
    \rho_i = \sum_{\substack{j=1 \\ j \neq i}}^N f (r_{ij}),
\end{equation}
where $f (r_{ij})$ represents the contribution to the electron density from atom $j$ at distance $r_{ij}$.

Various analytical forms for the pair potential, embedding energy, and electron density functions have been proposed in the literature \cite{daw_embedded-atom_1984,mishin_embedded-atom_2002,mishin_atomistic_2004_edited}.
Here we adopt the formulation proposed by Zhou et al. \cite{zhou_misfit-energy-increasing_2004}, in which
the pair potential is defined as:
\begin{equation}
    \label{eq:pair_potential}
    \phi(r) = \frac{A \exp[-\alpha (r/r_e - 1)]}{1 + (r/r_e - \kappa)^{20}} - \frac{B \exp[-\beta (r/r_e - 1)]}{1 + (r/r_e - \lambda)^{20}}.
\end{equation}
The first term represents pairwise repulsion and the second term accounts for pairwise attraction.
The electron density function has the same decaying exponential form as the attractive part of the pair potential, but with a different prefactor:
\begin{equation}
    \label{eq:density_function}
    f(r) = \frac{f_e \exp[-\beta (r/r_e - 1)]}{1 + (r/r_e - \lambda)^{20}}.
\end{equation}
The embedding function is defined as a piecewise function:
\begin{equation}
    \label{eq:embedding_function}
    F(\rho) =
    \begin{cases}
      \sum_{i=0}^3 F_{ni} \left( \frac{\rho}{\rho_n} - 1 \right)^i, & \rho < \rho_n, \rho_n = 0.85 \rho_e, \\
      \sum_{i=0}^3 F_i \left( \frac{\rho}{\rho_e} - 1 \right)^i, & \rho_n \leq \rho < \rho_0, \rho_0 = 1.15 \rho_e, \\
      F_e \left[ 1 - \log \left( \frac{\rho}{\rho_s} \right)^\eta \right] \left( \frac{\rho}{\rho_s} \right)^\eta, & \rho_0 \leq \rho.
    \end{cases}
\end{equation}

The functions $\phi(r)$, $f(r)$ and $F(\rho)$ contain in total 20 numerical parameters.
However, in optimizing our IPs, we allow variations only in a subset of seven parameters: $r_e$, $\beta$, $A$, $B$, $\kappa$, $\eta$, and $F_e$.
Three additional parameters ($\rho_s$, $\alpha$, $\lambda$) are fixed to account for gauge invariance in the EAM formalism \cite{daw_embedded-atom_1984}, and one parameter ($\rho_e$) is fixed because its variation either does not measurably affect simulated plastic strength or induces spurious phase transformations irrelevant to this study.
Finally, the remaining parameters are adjusted in response to any variations of the seven master parameters to ensure continuity up to the second derivatives of the pair, embedding, and electron density functions.
Physical constraints require all seven master parameters to remain non-negative, as:
\begin{itemize}
    \item $r_e$ is a distance scaling factor,
    \item $\beta$ defines the decay rate of interactions with increasing atomic separation,
    \item $A$, $B$, and $F_e$ are energy scaling factors,
    \item $\kappa$ is a dimensionless interaction cutoff parameter, and
    \item $\eta$ is an exponent that ensures proper behavior of the embedding function at large electron densities.
\end{itemize}
To account for different physical units and magnitudes of EAM potential parameters, and to improve numerical conditioning during IP training, we reparameterize the three functions defining the total energy in terms of logarithms of their parameters.

\subsection{Loss function}
\label{sec:loss-function}

A statistical model is typically trained by minimizing its prediction loss on reference data.
Although other types of GT data can be utilized in training an IP, total energies and forces on atoms computed over many atomic configurations are often used \cite{ercolessi_interatomic_1994,wen_kliff_2022}.
A parametric IP model is fitted to data by minimizing the loss function
\begin{equation}
    \label{eq:loss}
    L(\theta) = L_{\text{data}}(\theta) + \gamma L_{\text{reg}}(\theta),
\end{equation}
where $\theta$ denotes the parameters in the chosen model parameterization (logarithms of the seven tunable parameters in our case).
Here, the hyperparameter $\gamma$ defines a balance between closely fitting the reference data (the first loss term) and regularization penalty (the second loss term).
Although not always used or needed, the regularization penalty helps to mitigate data overfitting, especially dangerous when the model is fitted to a small dataset.
The same penalty prevents the model parameters from drifting toward unphysical values (e.g., diverging to infinity) that may nonetheless closely fit the training data.

Data contribution to the loss is most often defined as a weighted sum of squared residuals (i.e., prediction errors) computed over configuration energies and atomic forces,
\begin{equation}
    \label{eq:loss_data}
    L_{\text{data}}(\theta) = \sum_{m=1}^M  \left[ w_m^E (E_m^{\text{GT}} - E_m(\theta))^2 + \sum_{n_m=1}^{N_m} w_{n_m}^F \sum_{i=1}^3  (F_{n_m, i}^{\text{GT}} - F_{n_m, i}(\theta))^2 \right],
\end{equation}
where the superscript ``GT'' denotes the ground truth values, index $m$ labels atomic configurations, $n_m$ indexes the atoms in configuration $m$, and  $i$ denotes the Cartesian component of the force vector.
The weights $w_m^E$ and $w_{n_m}^F$ control the relative contributions of each data point in the training set.
The regularization term is chosen to be the $\ell_2$-norm,
\begin{equation}
    \label{eq:loss_reg}
    L_{\text{reg}}(\theta) = \sum_{n=1}^N (\theta_n - \theta_n^0)^2,
\end{equation}
where $\theta^0$ is the center of the regularization.

In defining the data loss  $L_{\text{data}}$, the developer decides on a weighting scheme for the squared residuals in Eq.~(\ref{eq:loss_data}), which may be used to select data to include in training.
For example, one can train using only atomic forces by setting $w_m^E = 0$ for all configurations, effectively excluding energy contributions, or alternatively train using only energies by setting all force weights $w_{n_m}^F = 0$.
Additionally, while total energy is uniquely defined only per configuration \cite{yu2011energy}, thus necessitating a single weight per configuration energy, forces are naturally partitioned by atom, permitting more flexible weighting.
Different weights can be assigned to force residuals on each atom in a configuration or even to each of the three Cartesian components of the per-atom force.
A single weight can be also assigned to forces on all atoms within each configuration in the dataset, which is achieved by setting $w_{n_m}^F = w_m^F$ for all $n_m = 1, 2, \dots, N_m$ and Cartesian components $i= 1, 2, 3$.

As described in the next section, in the latter of the two case studies undertaken in this work, we explore several different weighting schemes.
A common approach is to assign equal weights to all residuals of the same data type in the training dataset.
Alternatively, greater weights can be assigned to data deemed more relevant based on the developer's understanding of intended uses of the IP.
Instead, our IM method largely removes intuition from IP fitting by selecting (compressively) a minimal subset of reference data needed to constrain prediction uncertainty of the target QoIs and assigning non-arbitrary optimal weights for each retained data entry.

\subsection{Dataset}
\label{sec:dataset}

In this study, we train EAM potentials for Ta using three candidate datasets---referred to in the following as the ``MD--EAM-proxy'', ``MD--SNAP-proxy'', and ``DFT-reference'' datasets---which differ in the origin and number of included atomic configurations and/or in the method used to compute the reference training data.
The datasets are selected to facilitate exploration and step-by-step analysis of the utility and performance of the IM method, with each dataset introducing additional complexity and illuminating distinct challenges in reaching our overall objective.

The MD--EAM-proxy and MD--SNAP-proxy candidate datasets contain local atomic environments sampled from a single snapshot of an MD simulation of Ta crystal plasticity.
Given the large number of atoms in the snapshot---around 33 million---we screened all atomic environments in the snapshot to control the cost of IM iteration and retained 2,000 of them, selected for maximal diversity while also making sure that the selected subset is representative of the entire 33-million-atom snapshot.
To achieve this, we mapped all the environments onto the QUESTS descriptor space \cite{schwalbe-koda_model-free_2025} and used farthest-point sampling (FPS).
To efficiently perform this computation without computing a distance matrix for 33 million environments, we split the 33 million vector representations into approximately 3,300 chunks of 10,000 and used FPS to sample 2,000 environments from each chunk.
The sampled environments were then joined and partitioned again, forming new chunks of size 10,000 and allowing the process to be repeated.
The final 2,000 atomic environments were obtained after six iterations of this loop.
Furthermore, because these datasets consist of extracted local environments rather than complete atomic configurations, only the central atom force is used as GT data.
The forces on the other atoms in each extracted environment, as well as the total energy of the extracted environment, are not used because they depend on interactions with atoms outside the local environment and are therefore not well defined for the truncated configuration.

Ground truth forces on the central atom of each environment for these two datasets were then generated using two widely used IPs previously developed for Ta: the EAM potential reported in [\citenum{zhou_misfit-energy-increasing_2004}] and the SNAP potential in [\citenum{thompson_spectral_2015}].
In addition to greatly reducing the cost of generating GT data, using these two computationally efficient IPs as proxy sources of GT data permits direct validation of our AL workflow.
By comparing Ta strength predicted using an IP fitted to proxy GT data to strength predicted by the proxy potentials themselves, we can assess if and how well our IM method works toward its ultimate purpose---accurate prediction of Ta strength.
In comparison, similar direct validation against DFT is far beyond reach.

The DFT-reference candidate dataset is taken from the DFT training set used by \citet{sharma_development_2023} for developing an IP for the W--Ta system.
Here, we use only the Ta configurations from this dataset, resulting in 136 atomic configurations ranging in size from 51 to 1,280 atoms.
These configurations span a diverse set of Ta environments relevant to potential development, including strained bulk crystals, point defects, grain boundaries, dislocations, surfaces, stacking faults, void-like structures, and high-temperature solid and liquid configurations.

Unlike the MD--EAM-proxy and MD--SNAP-proxy datasets, in which only the central atom forces are used as reference data, in the DFT-reference case we explore five different combinations of configuration energies and per-atom forces by properly selecting different weighting schemes in the loss function, as described in Sec.~\ref{sec:loss-function}.
In the ``DFT-E'' subcase, only configuration energies are used, with weights applied per configuration.
The ``DFT-F'' subcase uses only atomic forces, with the same weights assigned to all per-atom forces in the same configuration.
In the ``DFT-F$_{\mathrm{atom}}$'' subcase, again only atomic forces are used, but weights are assigned individually to each atom.
In the ``DFT-E+F'' subcase, weights are assigned per configuration to both energies and forces.
Finally, the ``DFT-E+F$_{\mathrm{atom}}$'' subcase includes both energies and forces, using per-configuration weights for energies and per-atom weights for forces.
These five subcases form a useful testbed to examine how different combinations of training data influence the resulting potential.

By its functional form and regardless of the values of its seven variable parameters, our trial EAM potential predicts zero per-atom force for any atomic environment in which the central atom is a center of inversion symmetry.
Such is the case for all atoms in half of the 136 configurations in the DFT-reference dataset.
Since such zero forces contain no information about the parameters of our trial potential, forces from these centrosymmetric configurations are not included in the loss function.

\subsection{Information-matching method}
\label{sec:information-matching}

At the core of our information-matching (IM) method lies the Fisher information matrix (FIM), which quantifies the amount of information that observed data provides about the parameters of a parametric statistical model.
The inverse of the FIM provides a theoretical lower bound on the covariance of an unbiased estimator of the parameters, known as the Cram\'{e}r-Rao bound \cite{cramer1946,Rao1992}.
Mathematically, the FIM is defined as statistical expectation of the Hessian of the log-likelihood with respect to the model parameters,
\begin{equation}
    \label{eq:fim}
    \CalI_{ij}(\bfth) = -\mathbb{E} \left[ \frac{\partial^2}{\partial \theta_i \theta_j} \log p(x \mid \bfth) \right],
\end{equation}
where $p(x \mid \bfth)$ is the likelihood of data $x$ given parameters $\bfth$.
When multiple independent data points $\bfx = \{x_m\}_{m=1}^M$ with associated measurement uncertainties $\{\sigma_m\}$ are available, the total FIM equals the weighted sum of the individual per-data FIMs,
\begin{equation}
    \label{eq:fim_data}
    \CalI(\bfth) = \sum_{m=1}^M w_m \CalI_m(\bfth),
\end{equation}
where $\CalI_m$ is the FIM for data $x_m$ and $w_m = 1/\sigma_m^2$.
This relation emphasizes that the information contributed by each data point accumulates additively to yield the total expected information.

Importantly, the FIM is independent of the exact GT and can be computed prior to data acquisition.
This property enables its use in planning which data points would most effectively reduce parameter uncertainty---a critical consideration in fields where data acquisition is expensive, such as astrophysics \cite{vallisneri2008use}.
Consequently, the FIM underpins optimal experimental design by identifying data acquisition conditions that, once acquired, would maximally improve the precision of model predictions.
Beyond experimental design, the same property allows the FIM to be evaluated for downstream predictions of target quantities of interest (QoIs).
In this setting, the uncertainty measures for the target QoIs (denoted by $\delta_n$) specify desired precision thresholds.
In such application, the FIM quantifies the information about model parameters required to achieve predictions within these target uncertainties.

The IM method exploits both consequences of the GT-independent nature of the FIM and focuss specifically on \textbf{prediction uncertainty of one or more QoIs}, rather than on global parameter uncertainty.
In this work, IM enables the development of bespoke IPs tailored for accurately predicting metal strength in atomistic simulations.
The central objective is to identify a minimal set of training data such that, once acquired and used to fit the IP, the resulting parameter uncertainties are sufficiently bounded to meet the required precision of the target QoIs.
This condition is achieved by requiring that the information contained in the selected data exceeds the information needed to achieve the desired QoI uncertainties.
Rather than relying on simple scalar measures of gross information content, such as the determinant (D-optimality), IM is defined through a matrix inequality that accounts for the geometric alignment between two FIMs in parameter space.
Specifically, we require that the FIM summed over the data is greater than the FIM associated with the QoIs:
\begin{equation}
    \label{eq:matrix_inequality}
    \sum_{m \in \text{data}} w_m \CalI_m(\bfth) \succeq \CalJ(\bfth),
    \quad \CalJ(\bfth) = \sum_{n \in \text{QoI}} \frac{1}{\delta_n^2} \CalJ_n(\bfth) ,
\end{equation}
where $\CalJ_n$ denotes the FIM associated with QoI $n$.
The matrix inequality means that the difference between the left-hand side and the right-hand side is positive semi-definite.
This condition formalizes the information constraint and ensures that the optimal training data provide at least as much information as required to achieve the desired uncertainty of the target QoIs.

Finding a minimal informative subset is essential for reducing the cost of GT data acquisition and therefore constitutes the objective of data selection.
Given a pool of candidate data, compressive data selection within the IM method can be performed in several ways.
If the uncertainty for each data point is known, binary programming can be used to determine a minimal informative subset.
If uncertainties are unknown but can be bounded, one may instead optimize the data weights subject to an appropriate upper-bound constraint.
In this work, we follow the IM approach developed in \cite{kurniawan_information-matching_2026} and assume that data uncertainties are undefined, and perform \textbf{$\ell_1$-norm minimization} over the data weights:
\begin{equation}
    \label{eq:information-matching}
    \begin{aligned}
	& \text{minimize} && \sum_{m=1}^M w_m \\
	& \text{subject to} && w_m \geq 0, \\
	& && \sum_{m \in \text{data}} w_m \CalI_m(\bfth) \succeq \CalJ(\bfth),
    \end{aligned}
\end{equation}
where the constraint $w_m \geq 0$ preserves the statistical interpretation of the weights as inverse variances of the data.
The other aforementioned approaches for weight optimization can be incorporated into this framework by adding simple constraints, such as $w_m \in \{0,1\}$ or $w_m \leq w_{\max}$.

While other approaches have been proposed for selecting optimal training weights in IP development \cite{lenosky_highly_1997}, the IM framework provides a principled, information-theoretic basis for doing so.
Furthermore, when GT data have not yet been acquired, the optimal weights offer quantitative guidance on the target precision needed for collecting GT values.
This approach ensures a parsimonious selection of data minimally sufficient to achieve the desired precision in the prediction of the target QoIs.

\begin{figure}[!hbt]
    \centering
    \includegraphics[width=0.9\textwidth]{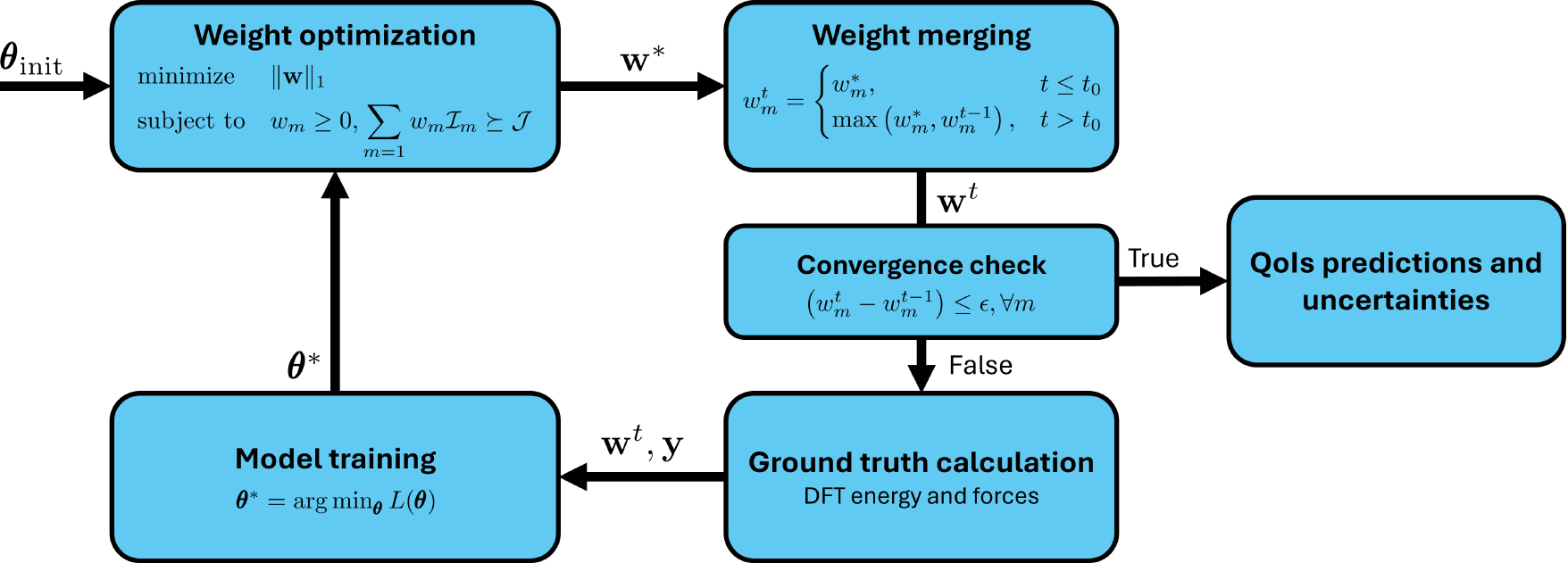}
    \caption[Workflow of active learning framework via information-matching]{
	Workflow of Active Learning by Information-Matching (ALIM).
	At each iteration, weights are optimized and merged across iterations.
	Then, the GT data are generated for those configurations with nonzero weights.
	The potential is then retrained using the generated GT data and their corresponding optimal weights.
	The loop continues until weight convergence, after which the optimized potential is used to predict quantities of interest and estimate their uncertainties.
    }
    \label{fig:information-matching_al}
\end{figure}

The IM method can be naturally integrated into an AL loop, forming what we refer to as the Active Learning by Information-Matching (ALIM) workflow.
The loop begins with an initial parameter guess $\bfth_{\text{init}}$, followed by weight optimization through solving Eq.~(\ref{eq:information-matching}).
The resulting weights $\bfw^*$ are merged with those from previous iterations.
For configurations with nonzero weights, GT data are computed if not already available, with the target uncertainty of each calculation set by the inverse square-root of its optimal weight.
These data, along with their optimal weights, are then used to retrain the model to get updated parameters $\bfth$ that serve as the starting point for the next iteration.
The procedure is repeated until the weights in successive iterations converge within a specified tolerance $\epsilon$.
At the conclusion of the AL loop, the resulting optimal potential can be used to predict the target QoIs and to estimate their corresponding prediction uncertainties.
For example, the prediction uncertainty of QoI $g_n$ can be estimated as
\begin{equation}
    \label{eq:preds_uncert}
    \sigma_{g_n} = \sqrt{\sum_{i, j} \left[\CalI^{-1} \right]_{ij} \frac{\partial g_n}{\partial \theta_i} \frac{\partial g_n}{\partial \theta_j}},
\end{equation}
where $\sigma_{g_n}$ is the estimated uncertainty of $g_n$, $\CalI$ is the FIM computed for the final optimal dataset with its optimized weights, and both $\CalI$ and the QoI derivatives are evaluated at converged optimal parameters $\bfth_{\text{opt}}$.
The ALIM workflow is illustrated in Fig.~\ref{fig:information-matching_al}.

Weight merging in the AL loop above is performed to prevent iterations from getting stuck in a limit cycle, i.e., oscillating between the same sets of weights and parameters.
A practical strategy is to compare the newly optimized weight $w_m^*$ with the weight from the previous iteration, $w_m^{t-1}$, and retain the larger value for each configuration, thus effectively preventing loss of information over subsequent iterations.
On the other hand, merging can result in disproportionately large weights assigned to some data, which may in turn over-constrain the model parameters, particularly when iterations start from a poor initial parameter guess.
To mitigate this issue, we run 5--10 initial iterations without weight merging and begin to collect and merge the weights only after the loop passes this initial burn-in stage.
The burn-in can be interpreted as giving the model time to explore data weights freely before enforcing memory of past iterations, leading to a  well-conditioned learning dynamic.

\subsection{Target material properties}
\label{sec:material-properties}

Although the main objective of this study is to develop an IP tailored for predicting the plastic strength of Ta, directly targeting plastic strength as a QoI for ALIM is prohibitively expensive, primarily due to the computational cost of evaluating the FIM for strength, which requires at least 15 large-scale MD simulations per iteration in the AL loop.
Prior work has demonstrated that several intermediate-scale properties (requiring only $\sim 100$ atoms to compute) correlate strongly with plastic strength in FCC single crystals \cite{jasperson_cross-scale_2025}.
Anticipating that similar trends may also hold for Ta, we adopt an indirect approach in which five intermediate-scale properties of BCC Ta, referred to here as indicator properties, are used as QoIs for ALIM rather than directly targeting strength. \textcolor{black}{We selected the five indicator properties from among easy to compute verification tests for new interatomic potentials implemented in the OpenKIM project.  Although these five indicators do not explicitly include quantities directly related to plastic strength, e.g. the Peierls stress, our prior analysis showed that they are strongly correlated with plastic strength and therefore provide a practical surrogate set that captures much of the relevant information.}

The selected indicator properties are the lattice constant, cohesive energy \cite{karls_equilibrium_2019,karls_equilibrium_2019-1}, and the elastic constants $c_{11}$, $c_{12}$, and $c_{44}$ \cite{huntington_elastic_1958,li_elastic_2019,tadmor_elastic_2019} for the cubic BCC Ta crystal, all computed using property tests implemented in OpenKIM \cite{OpenKIM-TD:475411767977:007,OpenKIM-TE:914032759339:007,OpenKIM-TE:391736780667:006a,OpenKIM-TD:011862047401:006,OpenKIM-TE:391736780667:006}.
At the end of the optimization, we assess \emph{post factum} whether these indicator properties contain sufficient information to enable reliable large-scale predictions of plastic strength.
While this step still necessitates evaluating the FIM for plastic strength, it reduces the computational cost to just one expensive strength FIM evaluation for the entire iterative process. \textcolor{black}{The ultimate QoI, the plastic strength of a body-centered cubic single crystal, was computed in large-scale MD simulations with the LAMMPS code \cite{thompson2022lammps}, following methods previously described in \cite{zepeda-ruiz_probing_2017,zepeda2021atomistic,bertin_crystal_2023}.  Additional details pertaining to MD simulations with perturbed trial EAM potentials are given in ‘Additional file 1’.}

For the target uncertainties $\delta_n$ of the indicator properties, an ideal choice would be values derived from the actual covariance between plastic strength and the indicator properties.
However, such covariance information for BCC metals is unknown, and selecting $\delta_n$ arbitrarily would be \emph{ad hoc}.
As a practical alternative, we instead compute $\delta_n$ using the prediction variance of the indicator properties over an ensemble of IPs for Ta available in OpenKIM \cite{Tadmor_Elliott_Sethna_Miller_Becker_2011}.
Furthermore, although the full covariance matrix could in principle be included in the QoI FIM calculation, the cross-correlations among the indicator properties are also unknown, and addressing them lies beyond the scope of this study.
The resulting target uncertainty for each indicator property is listed in Table~\ref{tab:target-uncert}.

\begin{table}[!hbt]
    \centering
    \begin{tabular}{M{5em} M{5em} M{5em} M{5em} M{5em}}
      \hhline{=====}
      $a_0$ (\AA) & $E_{\text{coh}}$ (eV) & $c_{11}$ (GPa) & $c_{12}$ (GPa) & $c_{44}$ (GPa) \\
      \hline
      $0.0075$ & $1.0869$ & $9.6143$ & $5.9992$ & $6.4602$ \\
      \hhline{=====}
    \end{tabular}
    \caption[Target uncertainties of the indicator properties]{
	Target uncertainties of the indicator properties, calculated from the standard deviations of the ensemble predictions over IPs for Ta in OpenKIM.
    }
    \label{tab:target-uncert}
\end{table}


\section{Results}
\label{sec:results}

\subsection{Training with proxy ground truth}
\label{sec:proxy_gt}

We use two previously developed IPs for Ta as proxy sources of GT data and compare the Ta strength predicted by our newly fitted (trial) potentials against predictions from these proxy potentials.
Even though such proxy GT data may not describe the plastic behavior of Ta at the highest accuracy, this comparison still serves as a validation of the ALIM workflow.
The two proxy datasets under study contain forces on central atoms from the same set of 2,000 diverse environments extracted from a large-scale MD simulation of Ta strength, as described in Sec.~\ref{sec:dataset}, and differ only by the method used to compute the forces: the MD--EAM-proxy dataset uses forces computed using the EAM potential~\cite{zhou_misfit-energy-increasing_2004}, while the MD--SNAP-proxy dataset uses forces computed using the SNAP potential~\cite{thompson_spectral_2015}.
The FIMs of the candidate environments are computed using the central-atom forces and are then matched to the FIM of five indicator properties: lattice constant, cohesive energy, and elastic constants $c_{11}$, $c_{12}$, and $c_{44}$.
After each iteration, the parameters $\bfth$ of the trial IPs are fitted by minimizing the loss function Eq.~(\ref{eq:loss}), evaluated only over the optimal (down-selected) environments with their corresponding optimal weights.
We refer to these two fitting cases according to the dataset used for training: the MD--EAM-proxy case and the MD--SNAP-proxy case.

\begin{figure}[!hbt]
    \centering
    \includegraphics[width=0.9\textwidth]{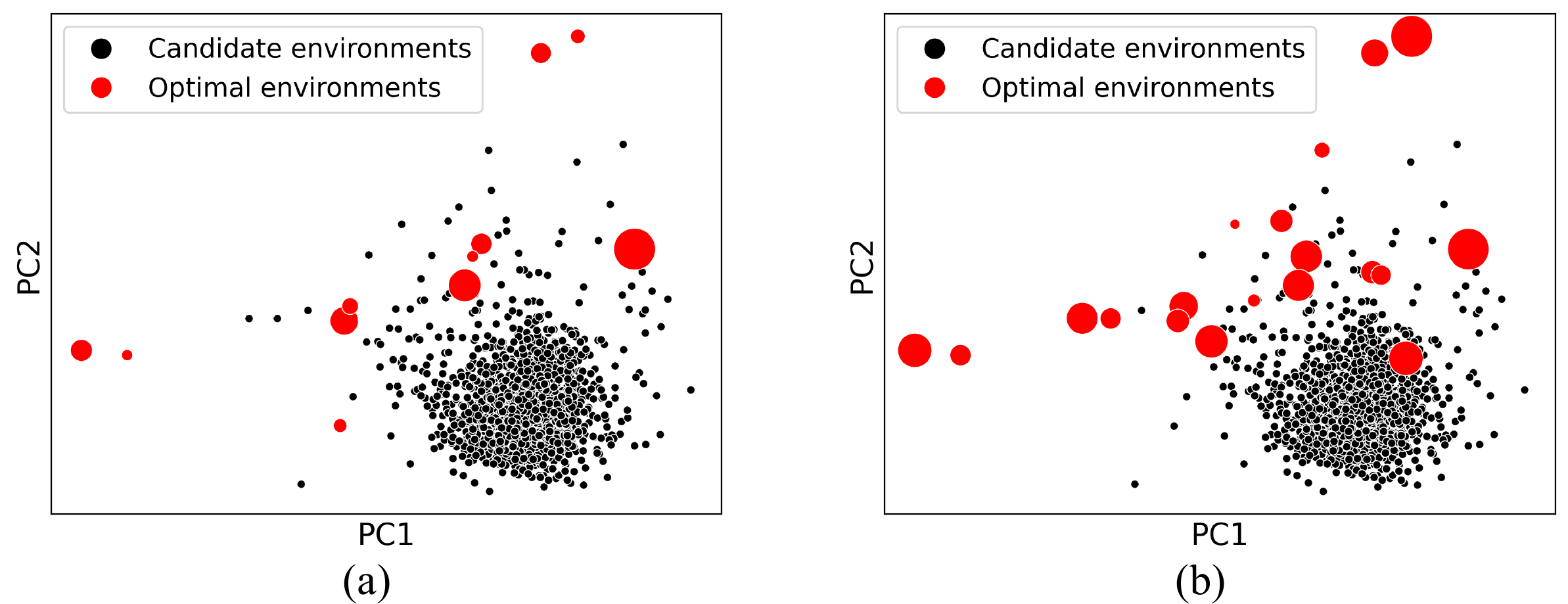}
    \caption[PCA representation of the selected optimal environments within the candidate dataset for cases with EAM and SNAP proxy ground truth]{
	PCA representation of the selected optimal environments for (a) MD--EAM-proxy and (b) MD--SNAP-proxy cases.
	The projection is computed from the QUESTS descriptors of the full candidate set.
	The candidate environments are shown as black dots, and selected environments as red circles with diameters proportional to the square root of their optimal weights (normalized within each set).
    }
    \label{fig:pca-eam-snap}
\end{figure}

Figure~\ref{fig:pca-eam-snap} shows the distribution of the candidate environments in the space of QUESTS descriptors~\cite{schwalbe-koda_model-free_2025}, projected onto the first two principal components.
Optimal environments selected in fully converged ALIM iterations are shown as red circles, with diameters proportional to the square root of their optimal weights, visually indicating their relative importance in fitting the two optimal trial IPs.
In both cases, as few as 0.5--1.0\% of the 2,000 candidate environments are sufficient to satisfy the information constraint and achieve the desired prediction precision for the five indicator properties.
Interestingly, the optimal environments identified as particularly informative for the indicator properties tend to lie in relatively sparsely populated regions of the projected QUESTS descriptor space.
However, this behavior should not be interpreted as a general preference for low-density regions; rather, it reflects the fact that ALIM selects environments based on their contribution to the QoI-relevant FIM, not on their representativeness in descriptor space.
Thus, candidate environments in the dense central cluster may provide largely redundant information for the selected indicator properties, whereas some peripheral environments supply complementary information needed to satisfy the information constraint.
Inspection of the selected environments, provided with the data repository, suggests that several of these peripheral configurations correspond to atoms near vacancies and dislocation cores.
Such environments break the local symmetry and can induce distinct local force responses, making them especially useful for constraining QoI-relevant model directions.

The strictest tolerance on the accuracy of GT calculations required to match the QoI FIM---computed as the inverse square root of the minimum optimal weight---is 5.87~meV/\AA\ and 7.98~meV/\AA\ for the MD--EAM-proxy and MD--SNAP-proxy cases, respectively.
While these tolerances have limited meaning for proxy GT forces, which contain no calculation uncertainty, they nevertheless fall within the typical accuracy range of DFT force calculations~\cite{kresse1996efficient}.

After solving Eq.~(\ref{eq:information-matching}) for the weights, each trial IP is trained on its respective GT data by fitting forces on the central atoms of the selected environments, with the optimal weights used in the loss function Eq.~(\ref{eq:loss}).
In the MD--EAM-proxy case, the reference and trial potentials share the same functional form, so explicit regularization is unnecessary.
In contrast, the SNAP proxy potential has a different functional form, requiring a fitting regularization to prevent the trial EAM parameters from diverging toward unphysical values (e.g., vanishing or diverging to infinity).
In this case, an initially large regularization hyperparameter $\lambda$ was gradually reduced to a value just above the threshold where parameter runaway was observed, i.e., where fitted parameter values diverged toward either zero or infinity.
The optimal parameters for each case are listed in Table~\ref{tab:opt-params-eam-snap}, along with the mean and standard deviation of dimensionless weighted residuals (w-residuals) computed over the training data to assess overall fit quality.

\begin{table}[!hbt]
    \centering
    \begin{tabular}{m{9em}<\raggedright r@{.}l r@{.}l}
	\hhline{=====}
      	\multirow[m]{2}{*}{\makecell[l]{\textbf{Model}\\\textbf{parameters}}}
	& \multicolumn{4}{c}{\rule{0pt}{2.5ex} \textbf{Dataset}} \\
        \cmidrule(lr){2-5}
        & \multicolumn{2}{c}{\textbf{MD--EAM-proxy}} & \multicolumn{2}{c}{\textbf{MD--SNAP-proxy}} \\
	\hline
	$r_e$ (\AA) & \phantom{000}$ 2$ & $8559$ & \phantom{000}$ 2$ & $8682$ \\
	$\beta$     & \phantom{000}$ 4$ & $5269$ & \phantom{000}$ 4$ & $7375$ \\
	$A$ (eV)    & \phantom{000}$ 0$ & $6197$ & \phantom{000}$ 0$ & $7204$ \\
	$B$ (eV)    & \phantom{000}$ 1$ & $0508$ & \phantom{000}$ 1$ & $0783$ \\
	$\kappa$    & \phantom{000}$ 0$ & $1782$ & \phantom{000}$ 0$ & $2059$ \\
	$\eta$      & \phantom{000}$ 0$ & $7458$ & \phantom{000}$ 0$ & $5523$ \\
	$F_e$ (eV)  & \phantom{000}$-5$ & $1358$ & \phantom{000}$-5$ & $7449$ \\
	\hline
	Mean w-residuals   & $0$ & $0001$ & $8$ & $7942$ \\
	Stdev. w-residuals & $0$ & $0013$ & $8$ & $6157$ \\
	\hhline{=====}
    \end{tabular}
    \caption[Optimal EAM parameters obtained for cases fitted to EAM and SNAP ground truth]{
	Optimal EAM parameters obtained using the ALIM method, fitted to MD--EAM-proxy and MD--SNAP-proxy GT.
	Units are given in parentheses; parameters without units are dimensionless.
	The mean and standard deviation of the weighted residuals are reported as measures of fit quality.
    }
    \label{tab:opt-params-eam-snap}
\end{table}

Assuming Gaussian noise---implicit in the least-squares loss---a well-specified model would produce weighted residuals with a mean near zero and a standard deviation close to or below one.
This is indeed observed for the MD--EAM-proxy case, as expected, since the fitted model and GT share the same functional form.
In contrast, the MD--SNAP-proxy case exhibits both a large mean and a large standard deviation of the weighted residuals, indicating significant \emph{model error}.

The large model error for the MD--SNAP-proxy case arises from at least two sources.
From a model-form perspective, the SNAP proxy GT is generated by a more flexible potential form than the fitted EAM model.
As a result, the fitted EAM potential may not have sufficient flexibility to reproduce all features of the SNAP GT.
Additionally, there is a mismatch in the absolute reference energy between the fitted EAM potential and the SNAP proxy considered in this study.
Because the EAM potential is fitted only to forces from condensed-phase configurations, the loss function constrains only the local shape of the SNAP potential energy surface near those configurations, but not the absolute energy difference between the condensed phase and the isolated-atom limit.
As a result, the absolute energy scale cannot be reliably learned during fitting.

To further illustrate the effect of model error, we use the trained IPs to compute the five indicator properties as well as the ultimate target QoI of plastic strength, and evaluate the prediction errors relative to GT values obtained from the two proxy GT potentials.
The results, along with the associated uncertainties computed using Eq.~(\ref{eq:preds_uncert}), are listed in Table~\ref{tab:preds-uncert-eam-snap}.
For the MD--EAM-proxy case, the prediction uncertainties closely match the actual errors, indicating that the uncertainties reliably reflect prediction accuracy.
In contrast, for the MD--SNAP-proxy case, while the prediction uncertainties satisfy the uncertainty constraints for the indicator properties (cf. Table~\ref{tab:target-uncert}), they systematically underestimate the true errors and give overconfident estimates of prediction accuracy.

\begin{table}[!hbt]
    \centering
    \begin{tabular}{m{3.5em} p{10em}<{\raggedright} r@{.}l r@{.}l}
      \hhline{======}
      \multicolumn{2}{l}{\multirow[m]{2}{*}{\textbf{Quantities of interest}}}
      & \multicolumn{4}{c}{\rule{0pt}{2.5ex}\textbf{Dataset}} \\
      \cmidrule(lr){3-6}
      \multicolumn{2}{l}{}
      & \multicolumn{2}{c}{\textbf{MD--EAM-proxy}}
      & \multicolumn{2}{c}{\textbf{MD--SNAP-proxy}} \\
      \hline
      \multirow[m]{4}{3.5em}{\centering\shortstack{$a_0$ \\ (\AA)}}
      & Mean
	& \phantom{000}$3$ & $3019$\phantom{00}
	& \phantom{000}$3$ & $3835$\phantom{00} \\
      & Error
	& \phantom{000}$ 0$ & $0007$\phantom{00}
	& \phantom{000}$-0$ & $0672$\phantom{00} \\
      & Uncertainty
	& \phantom{000}$0$ & $0053$\phantom{00}
	& \phantom{000}$0$ & $0034$\phantom{00} \\
      & Inflated uncertainty
	& \multicolumn{2}{c}{\phantom{000}--\phantom{00}}
	& \phantom{000}$0$ & $1183$\phantom{00} \\
      \hline
      \multirow[m]{4}{3.5em}{\centering\shortstack{$E_{\text{coh}}$ \\ (eV/atom)}}
      & Mean                 & $ 8$ & $1491$          & $ 8$ & $3938$ \\
      & Error                & $-0$ & $0591$          & $ 3$ & $4578$\tnote{a} \\
      & Uncertainty          & $ 0$ & $2301$          & $ 0$ & $2321$ \\
      & Inflated uncertainty & \multicolumn{2}{c}{--} & $ 8$ & $1531$ \\
      \hline
      \multirow[m]{4}{3.5em}{\centering\shortstack{$c_{11}$ \\ (GPa)}}
      & Mean                 & $253$ & $93  $         & $197$ & $17  $ \\
      & Error                & $  8$ & $7936$         & $ 73$ & $068 $ \\
      & Uncertainty          & $  6$ & $8172$         & $  4$ & $1493$ \\
      & Inflated uncertainty & \multicolumn{2}{c}{--} & $145$ & $77  $ \\
      \hline
      \multirow[m]{4}{3.5em}{\centering\shortstack{$c_{12}$ \\ (GPa)}}
      & Mean                 & $148$ & $99  $         & $126$ & $21  $ \\
      & Error                & $  8$ & $6992$         & $ 29$ & $669 $ \\
      & Uncertainty          & $  5$ & $7185$         & $  3$ & $8332$ \\
      & Inflated uncertainty & \multicolumn{2}{c}{--} & $134$ & $66  $ \\
      \hline
      \multirow[m]{4}{3.5em}{\centering\shortstack{$c_{44}$ \\ (GPa)}}
      & Mean                 & $82$ & $691 $          & $ 86$ & $872 $ \\
      & Error                & $-0$ & $7846$          & $-13$ & $152 $ \\
      & Uncertainty          & $ 3$ & $5366$          & $  1$ & $7173$ \\
      & Inflated uncertainty & \multicolumn{2}{c}{--} & $ 60$ & $329 $ \\
      \hline
      \multirow[m]{4}{3.5em}{\centering\shortstack{Strength \\ (GPa)}}
      & Mean                 & $2$ & $5299$           & $2$ & $0014$ \\
      & Error                & $0$ & $0035$           & $0$ & $1594$ \\
      & Uncertainty          & $0$ & $0466$           & $0$ & $0288$ \\
      & Inflated uncertainty & \multicolumn{2}{c}{--} & $1$ & $0108$ \\
      \hhline{======}
    \end{tabular}

    \begin{tablenotes}
	\footnotesize
	\item[a] Here, we instead use the potential energy prediction of the SNAP proxy as the GT value, i.e., no isolated-atom energy is subtracted.
	The SNAP proxy includes a constant energy shift that sets a different absolute energy reference.
	Comparing cohesive energies would therefore introduce a reference-energy mismatch.
    \end{tablenotes}

    \caption[Predictions, errors, and uncertainties of the target material properties for the MD--EAM-proxy and MD--SNAP-proxy cases]{
	Predicted values, errors, and uncertainties for the target material properties in the MD--EAM-proxy and MD--SNAP-proxy cases.
	The first five rows correspond to the indicator properties used in ALIM; the last row reports the plastic strength of Ta.
	Errors are defined as the difference between GT and prediction (negative values indicate overestimation).
	Uncertainties are computed using Eq.~(\ref{eq:preds_uncert}); inflated uncertainties are obtained by rescaling these propagated uncertainties by a model-error scale inferred from the fitting residuals.
    }
    \label{tab:preds-uncert-eam-snap}
\end{table}

This discrepancy highlights a fundamental limitation of FIM-based approaches: while they can reduce parameter uncertainty within a given model form, they do not indicate how the model itself should be improved to reduce prediction error.
When the model is insufficiently expressive or otherwise misspecified, the estimated uncertainties primarily reflect limitations of the training data rather than intrinsic model inadequacy, causing predictions to appear more confident than they truly are.
A practical remedy is to apply a \emph{post hoc} correction that inflates the propagated uncertainty estimates by a model-error scale inferred from the fitting residuals~\cite{frederiksen_bayesian_2004,pernot_parameter_2017,kurniawan_bayesian_2022}.
While this simple correction can mitigate overconfidence, it often substantially overestimates the prediction error, reducing the practical utility and reliability of the predictions.
A similar over-correction is observed in the plastic strength prediction for the MD--SNAP-proxy case: the uncorrected uncertainties (which do not account for model error) are roughly an order of magnitude smaller than the actual prediction error, while the corrected (inflated) uncertainties substantially exceed the prediction error, reaching up to 50\% of the predicted strength value itself.
In addition to revealing the inadequacy of the trial potential for fitting the SNAP GT data, these observations highlight the need for cautious interpretation of the corrected uncertainty estimates~\cite{pernot2017parameter}.

\subsection{Training with DFT reference data}
\label{sec:dft}

The second case study follows a common workflow in IP development, where the GT dataset consists of relatively small ($\sim$ 100 atom) configurations with energies and forces on atoms computed using accurate but computationally expensive DFT calculations.
We examine five subcases that use DFT-reference dataset, each defined by a different selection of quantities---energies and/or forces---included in the training loss and distinct weighting schemes.
Our goal is to assess whether, and to what extent, these weighting choices influence the resulting IP prediction uncertainties for the same five indicator properties and for the ultimate QoI, the plastic strength of Ta.
We also assess prediction errors for the indicator properties by comparing the IP predictions to their DFT GT values.
In contrast, an analogous validation for plastic strength is not feasible because computing strength directly from DFT is prohibitively expensive.

\begin{figure}[!hbt]
    \centering
    \includegraphics[width=0.9\textwidth]{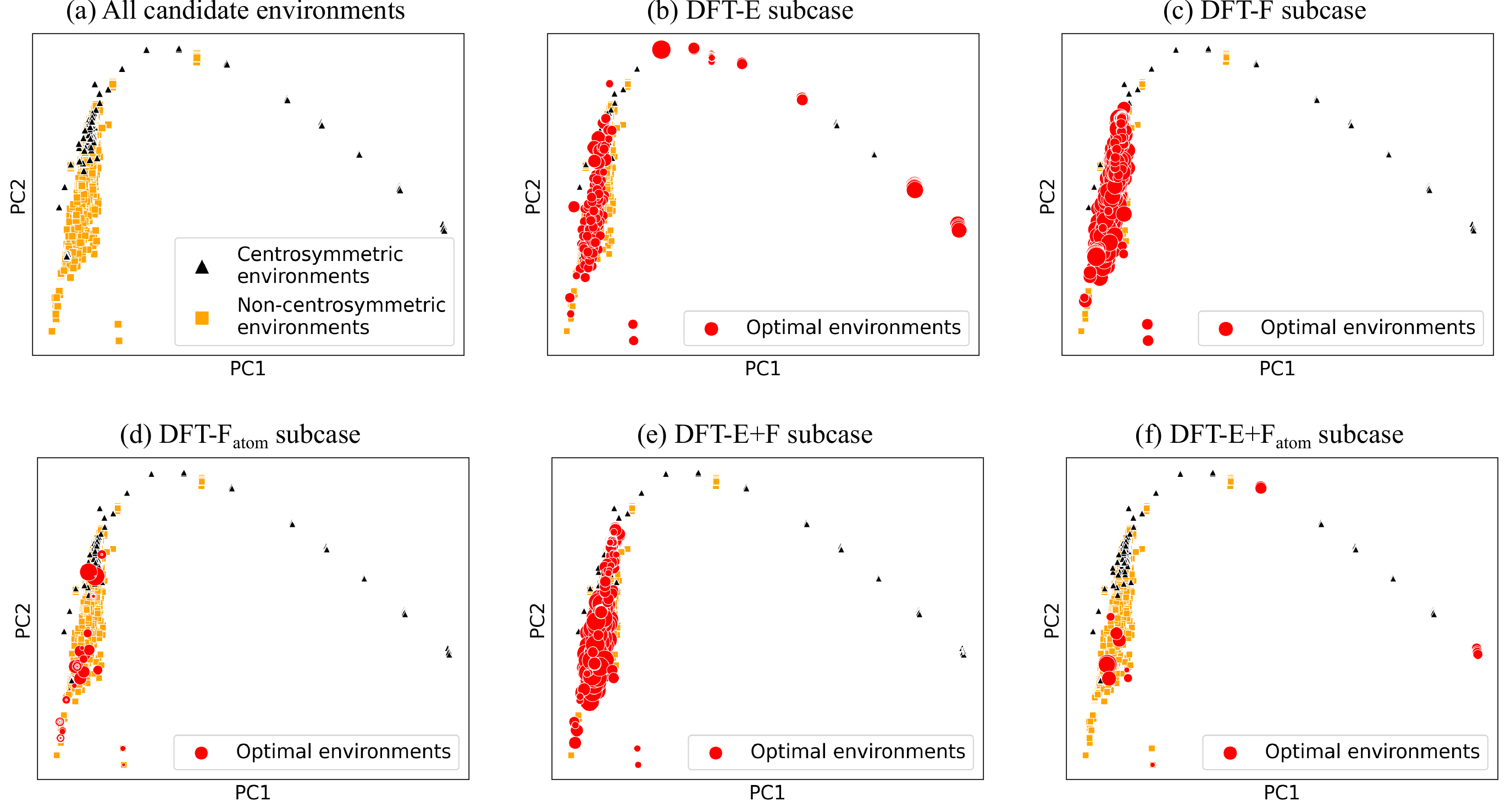}
    \caption[PCA representation of the optimal environments over the training data for DFT ground truth subcases]{
	PCA projection of the QUESTS descriptor space showing candidate environments and those selected for five ALIM subcases fitted to different subsets of the DFT-reference dataset.
	Panel (a) shows all candidates: black dots denote centrosymmetric environments (energy-only data), and orange dots denote non-centrosymmetric environments (energy and force data).
	Panels (b)--(f) show the selected environments for the DFT-E, DFT-F, DFT-F$_{\text{atom}}$, DFT-E+F, and DFT-E+F$_{\text{atom}}$ subcases.
	Red circles' diameters are proportional to the square root of the optimal weights, normalized within each subcase.
    }
    \label{fig:pca-dft}
\end{figure}

Figure~\ref{fig:pca-dft} shows all atomic environments in the dataset, along with the optimal environments selected at the completion of the ALIM iterations, projected onto the first two principal components of the QUESTS descriptor space.
To highlight differences in data coverage across the five subcases, each subfigure distinguishes centrosymmetric environments---where the force on the central atom is trivially zero and only the total energy can be used for fitting (black dots)---from the much larger set of non-centrosymmetric environments, for which force data are also available (orange dots).
Most candidate environments form a dense cluster, whereas the centrosymmetric environments span a broader region of the PCA subspace, extending into areas of descriptor space that are inaccessible when training solely on force information.

Table~\ref{tab:opt-weights-dft} lists the number of atomic configurations (not atomic environments) from which energy and/or force data were selected for fitting in each of the five subcases, along with the total number of energy and force data included in the optimal training set.
The next two rows report the minimum and maximum uncertainties permitted for DFT energy calculations---computed as the inverse square root of the minimum and maximum weights of the down-selected energy data---to match the FIM of the five indicator properties.
The bottom two rows list the corresponding minimum and maximum uncertainties required for DFT force calculations.
Although the number of down-selected atomic configurations is similar in magnitude across the five subcases, the number of selected environments in the subcases using per-atom weighting is much smaller than in those using per-configuration weighting (see the third row).
This occurs because, when permitted to assign weights to individual environments within a configuration, ALIM typically selects only a few highly informative environments, leaving most local environments unused (with zero weights).
In contrast, assigning the same nonzero weight to all per-atom forces in a selected configuration tends to redundantly sample less informative environments.
Thus, per-atom weighting allows the method to better prioritize the most informative environments while avoiding oversampling of redundant local structures.

\begin{table}[!hbt]
    \centering
    \resizebox{\textwidth}{!}{
      \begin{tabular}{l @{\hspace{2em}} r@{.}l r@{.}l r@{.}l r@{.}l r@{.}l}
	\hhline{===========}
	& \multicolumn{10}{c}{\rule{0pt}{2.5ex} \textbf{Dataset}} \\
	\cmidrule(lr){2-11}
	& \multicolumn{2}{c}{\makecell[c]{\textbf{DFT-E}}}
        & \multicolumn{2}{c}{\makecell[c]{\textbf{DFT-F}}}
        & \multicolumn{2}{c}{\makecell[c]{\textbf{DFT-F$_{\mathrm{atom}}$}}}
        & \multicolumn{2}{c}{\makecell[c]{\textbf{DFT-E+F}}}
        & \multicolumn{2}{c}{\makecell[c]{\textbf{DFT-E+F$_{\mathrm{atom}}$}}} \\
	\hline
	Number of configurations
	  & \multicolumn{2}{c}{$17$}
	  & \multicolumn{2}{c}{$7$}
	  & \multicolumn{2}{c}{$11$}
	  & \multicolumn{2}{c}{$7$}
	  & \multicolumn{2}{c}{$7$} \\
	\hline
	Number of energy data
	  & \multicolumn{2}{c}{$17$}
	  & \multicolumn{2}{c}{--}
	  & \multicolumn{2}{c}{--}
	  & \multicolumn{2}{c}{$1$}
	  & \multicolumn{2}{c}{$2$} \\
	Number of force data
	  & \multicolumn{2}{c}{--}
	  & \multicolumn{2}{c}{$5,235$}
	  & \multicolumn{2}{c}{$456$}
	  & \multicolumn{2}{c}{$3,744$}
	  & \multicolumn{2}{c}{$27$} \\
	\hline
	Min $\sigma_m^E$ (meV/atom)
	  & \phantom{000}$10$  & $011$\phantom{00}
	  & \multicolumn{2}{c}{--}
	  & \multicolumn{2}{c}{--}
	  & \phantom{000}$707$ & $80$\phantom{00}
	  & \phantom{000000}$ 90$ & $562$ \\
	Max $\sigma_m^E$ (meV/atom)
	  & $79$  & $674$
	  & \multicolumn{2}{c}{--}
	  & \multicolumn{2}{c}{--}
	  & $707$ & $80$
	  & $132$ & $22 $ \\
	Min $\sigma_{n_m}^F$ (meV/\AA/atom)
	  & \multicolumn{2}{c}{--}
	  & \phantom{000}$0$  & $9974$\phantom{00}
	  & \phantom{000}$19$ & $200$\phantom{00}
	  & \phantom{000}$0$  & $8656$\phantom{00}
	  & \phantom{000000}$34$ & $293$ \\
	Max $\sigma_{n_m}^F$ (meV/\AA/atom)
	  & \multicolumn{2}{c}{--}
	  & $    4$ & $6550$
	  & $3,047$ & $7$
	  & $    6$ & $7245$
	  & $  487$ & $54 $ \\
	\hhline{===========}
      \end{tabular}
    }
    \caption[Number of selected configurations with the minimum and maximum data uncertainties for the subcases fitted to DFT ground truth]{
	Number of selected atomic configurations, total selected energy and force data (counting each Cartesian force component separately), and the minimum and maximum required uncertainties for DFT GT energy and force calculations in each subcase.
	Uncertainties are computed as the inverse square root of the optimal weights and represent target error bounds for the DFT calculations.
	For subcases using per-configuration force weighting, the listed force uncertainties are divided by the number of atoms to yield per-atom target force error bounds.
    }
    \label{tab:opt-weights-dft}
\end{table}

For each of the five subcases of data weighting, the ALIM workflow selected an optimal subset of informative atomic configurations and/or environments together with their corresponding optimal weights, which were then used to fit five EAM potentials by minimizing the loss function in Eq.~(\ref{eq:loss}).
As in the MD--SNAP-proxy case, regularization was applied, with the regularization penalty parameter $\gamma$ tuned separately for each subcase.
The resulting optimal EAM parameters are listed in Table~\ref{tab:opt-params-dft}, along with the mean and standard deviation of the weighted residuals (w-residuals), which serve as indicators of fit quality in each subcase.
Predictions of the intermediate target QoIs obtained from the newly fitted IPs, together with their errors, uncertainties, and inflated uncertainties accounting for model error are presented in Table~\ref{tab:preds-uncert-dft}.
The bottom row of the same table lists the predicted plastic strength of Ta from all five IPs and the corresponding prediction uncertainties and inflated uncertainties.

\begin{table}[!hbt]
    \centering
    \resizebox{\textwidth}{!}{
	\begin{tabular}{m{9em}<\raggedright r@{.}l r@{.}l r@{.}l r@{.}l r@{.}l}
	  \hhline{===========}
	  \multirow[m]{2}{*}{\makecell[l]{\textbf{Model}\\\textbf{parameters}}}
	  & \multicolumn{10}{c}{\rule{0pt}{2.5ex} \textbf{DFT-reference subcases}} \\
	  \cmidrule(lr){2-11}
	  & \multicolumn{2}{c}{\makecell[c]{\textbf{DFT-E}}}
	  & \multicolumn{2}{c}{\makecell[c]{\textbf{DFT-F}}}
	  & \multicolumn{2}{c}{\makecell[c]{\textbf{DFT-F$_{\mathrm{atom}}$}}}
	  & \multicolumn{2}{c}{\makecell[c]{\textbf{DFT-E+F}}}
	  & \multicolumn{2}{c}{\makecell[c]{\textbf{DFT-E+F$_{\mathrm{atom}}$}}} \\
	  \hline
	  $r_e$ (\AA) & $ 2$ & $9416$ & $ 2$ & $4584$ & $ 2$ & $7044$ & $ 2$ & $4720$ & $ 2$ & $6797$ \\
	  $\beta$     & $ 4$ & $5703$ & $ 4$ & $6567$ & $ 4$ & $9898$ & $ 4$ & $6893$ & $ 4$ & $9102$ \\
	  $A$ (eV)    & $ 0$ & $6470$ & $ 1$ & $4576$ & $ 0$ & $7435$ & $ 1$ & $3775$ & $ 0$ & $8713$ \\
	  $B$ (eV)    & $ 1$ & $1770$ & $ 0$ & $5058$ & $ 0$ & $9156$ & $ 0$ & $3928$ & $ 1$ & $2900$ \\
	  $\kappa$    & $ 0$ & $1660$ & $ 0$ & $2632$ & $ 0$ & $1850$ & $ 0$ & $2664$ & $ 0$ & $1474$ \\
	  $\eta$      & $ 0$ & $5946$ & $ 0$ & $5248$ & $ 0$ & $7213$ & $ 0$ & $2747$ & $ 0$ & $5816$ \\
	  $F_e$ (eV)  & $-6$ & $0244$ & $-4$ & $1796$ & $-6$ & $6282$ & $-5$ & $4055$ & $-4$ & $9709$ \\
	  \hline
	  Mean w-residuals
	    & $-0$ & $8951$
	    & $ 8$ & $2463 \times 10^{-9}$
	    & $-0$ & $2082$
	    & $ 0$ & $0041$
	    & $-2$ & $7504$ \\

	  Stdev. w-residuals
	    & \phantom{000} $6$  & $0573$\phantom{000}
	    & \phantom{000} $0$  & $8962$\phantom{000}
	    & \phantom{000} $6$  & $5473$\phantom{000}
	    & \phantom{000} $1$  & $2627$\phantom{000}
	    & \phantom{0000}$23$ & $028 $ \\
	  \hhline{===========}
	\end{tabular}
    }
    \caption[Optimal EAM parameters for subcases fitted to DFT ground truth]{
	Optimal EAM parameters obtained using the ALIM method for each DFT subcase.
	Units are given in parentheses; parameters without units are dimensionless.
	The mean and standard deviation of the weighted residuals are reported as measures of fit quality.
    }
    \label{tab:opt-params-dft}
\end{table}

\begin{table}[!hbt]
    \centering
    \resizebox{\textwidth}{!}{
      \begin{tabular}{m{3.5em} p{9em}<{\raggedright} r@{.}l r@{.}l r@{.}l r@{.}l r@{.}l}
	\hhline{============}
	\multicolumn{2}{l}{\multirow[m]{2}{*}{\textbf{Quantities of interest}}}
	& \multicolumn{10}{c}{\rule{0pt}{2.5ex} \textbf{DFT-reference subcases}} \\
	\cmidrule(lr){3-12}
	\multicolumn{2}{l}{}
	& \multicolumn{2}{c}{\makecell[c]{\textbf{DFT-E}}}
	& \multicolumn{2}{c}{\makecell[c]{\textbf{DFT-F}}}
	& \multicolumn{2}{c}{\makecell[c]{\textbf{DFT-F$_{\mathrm{atom}}$}}}
	& \multicolumn{2}{c}{\makecell[c]{\textbf{DFT-E+F}}}
	& \multicolumn{2}{c}{\makecell[c]{\textbf{DFT-E+F$_{\mathrm{atom}}$}}} \\
	\hline

	\multirow[m]{4}{3.5em}{\centering\shortstack{$a_0$ \\ (\AA)}}
	& Mean
	  & \phantom{000}$3$ & $3536$\phantom{000}
	  & \phantom{000}$3$ & $3280$\phantom{000}
	  & \phantom{000}$3$ & $2865$\phantom{000}
	  & \phantom{000}$3$ & $3184$\phantom{000}
	  & \phantom{000}$3$ & $1296$\phantom{000} \\
	& Error
	  & $-0$ & $0346$
	  & $-0$ & $0090$
	  & $ 0$ & $0325$
	  & $ 0$ & $0006$
	  & $ 0$ & $1894$ \\
	& Uncertainty
          & $0$ & $0027$
	  & $0$ & $0033$
	  & $0$ & $0020$
	  & $0$ & $0023$
	  & $0$ & $0024$ \\
	& Inflated uncertainty
	  & $0$ & $0257$
	  & $0$ & $0803$
	  & $0$ & $1074$
	  & $0$ & $0676$
	  & $0$ & $1145$ \\
	\hline

	\multirow[m]{4}{3.5em}{\centering\shortstack{$E_{\text{coh}}$ \\ (eV/atom)}}
	& Mean
	  & $9$ & $6464$
	  & $3$ & $6177$
	  & $7$ & $9983$
	  & $4$ & $5802$
	  & $8$ & $2321$ \\
	& Error
	  & $-0$ & $0082$
	  & $ 6$ & $0204$
	  & $ 1$ & $6398$
	  & $ 5$ & $0579$
	  & $ 1$ & $4061$ \\
	& Uncertainty
          & $0$ & $0232$
	  & $0$ & $0801$
	  & $0$ & $0963$
	  & $0$ & $0682$
	  & $0$ & $1061$ \\
	& Inflated uncertainty
	  & $0$ & $2216$
	  & $1$ & $9638$
	  & $5$ & $0906$
	  & $1$ & $9926$
	  & $5$ & $0078$ \\
	\hline

	\multirow[m]{4}{3.5em}{\centering\shortstack{ $c_{11}$ \\ (GPa)}}
	& Mean
	  & $255$ & $24$
	  & $190$ & $48$
	  & $337$ & $12$
	  & $237$ & $98$
	  & $492$ & $95$  \\
	& Error
	  & $  14$ & $755$
	  & $  79$ & $516$
	  & $ -67$ & $125$
	  & $  32$ & $020$
	  & $-222$ & $95 $ \\
	& Uncertainty
          & $5$ & $4122$
	  & $4$ & $9561$
	  & $2$ & $3834$
	  & $3$ & $9663$
	  & $7$ & $0384$ \\
	& Inflated uncertainty
	  & $ 51$ & $644$
	  & $121$ & $46	$
	  & $126$ & $01	$
	  & $115$ & $84	$
	  & $332$ & $24 $ \\
	\hline

	\multirow[m]{4}{3.5em}{\centering\shortstack{$c_{12}$ \\ (GPa)}}
	& Mean
	  & $140$ & $84$
	  & $165$ & $05$
	  & $237$ & $83$
	  & $203$ & $46$
	  & $221$ & $86$ \\
	& Error
	  & $ 15$ & $155 $
	  & $ -9$ & $0351$
	  & $-81$ & $826 $
	  & $-47$ & $458 $
	  & $-65$ & $855 $ \\
	& Uncertainty
          & $1$ & $8387$
	  & $2$ & $8316$
	  & $2$ & $0336$
	  & $2$ & $2591$
	  & $3$ & $1113$   \\
	& Inflated uncertainty
	  & $ 17$ & $546$
	  & $ 69$ & $397$
	  & $107$ & $52	$
	  & $ 65$ & $982$
	  & $146$ & $87 $ \\
	\hline

	\multirow[m]{4}{3.5em}{\centering\shortstack{$c_{44}$ \\ (GPa)}}
	& Mean
	  & $ 93$ & $191$
	  & $ 66$ & $418$
	  & $ 82$ & $725$
	  & $ 64$ & $636$
	  & $163$ & $38 $ \\
	& Error
	  & $-19$ & $191 $
	  & $  7$ & $5819$
	  & $ -8$ & $7247$
	  & $  9$ & $3644$
	  & $-89$ & $383 $ \\
	& Uncertainty
          & $2$ & $5785$
	  & $1$ & $3964$
	  & $0$ & $7806$
	  & $0$ & $5831$
	  & $1$ & $4290$ \\
	& Inflated uncertainty
	  & $24$ & $604$
	  & $34$ & $222$
	  & $41$ & $272$
	  & $17$ & $032$
	  & $67$ & $456$ \\
	\hline

	\multirow[m]{3}{3.5em}{\centering\shortstack{Strength \\ (GPa)}}
	& Mean
	  & $2$ & $9730$
	  & $1$ & $4530$
	  & $1$ & $9797$
	  & $1$ & $3771$
	  & $5$ & $4042$ \\
	& Uncertainty
          & $0$ & $0821$
	  & $0$ & $1803$
	  & $0$ & $0157$
	  & $0$ & $0342$
	  & $0$ & $1536$ \\
	& Inflated uncertainty
	  & $0$ & $7833$
	  & $4$ & $4183$
	  & $0$ & $8308$
	  & $0$ & $9999$
	  & $7$ & $2515$ \\
	\hhline{============}
      \end{tabular}
    }
    \caption[Predictions and uncertainties of the target material properties from potentials fitted to DFT ground truth]{
	Predictions, errors, uncertainties, and inflated uncertainties of the target material properties obtained from EAM potentials fitted to DFT GT for each subcase.
	Results are reported for the five indicator properties used in the ALIM method and the large-scale target plastic strength of Ta.
        Errors are defined as the difference between GT and prediction (negative values indicate overestimation).
        Uncertainties are computed using Eq.~(\ref{eq:preds_uncert}); inflated uncertainties are obtained by rescaling these propagated uncertainties by a model-error scale inferred from the fitting residuals, providing a basis for assessing reliability when ground truth data are unavailable (e.g., plastic strength).
    }
    \label{tab:preds-uncert-dft}
\end{table}

For these subcases, although the mean and standard deviation of the residuals suggest that the optimal EAM potentials fit the training data reasonably well, they alone do not imply that the model error uncertainty is negligible.
In fact, the model errors for these subcases are substantial, such that there are many instances in Table~\ref{tab:preds-uncert-dft} where the indicator property predictions are overconfident.
Applying the suggested uncertainty inflation correction improves the agreement between predicted uncertainties and actual errors in most instances.
Nonetheless, the corrected uncertainties sometimes significantly overestimate the true errors, such as for subcases involving per-atom force weighting, thus call for cautious interpretation of the inflated uncertainties.
For the plastic strength predictions, direct comparison with GT is not possible, since no reference DFT data is available.
Instead, based on analyses in previous cases with proxy GT, we anticipate that the inflated uncertainty serves as a conservative bound on the true error.

Comparing the errors of the indicator properties with the strength predictions in Table~\ref{tab:preds-uncert-dft}, we observe an apparent positive correlation between the plastic strength and the elastic constants $c_{11}$ and $c_{44}$, and a negative correlation with the lattice constant $a_0$.
The inflated uncertainties for these three indicator properties also appear to correlate with the uncertainty in strength, though the correlation is weaker.
These trends are qualitatively consistent with findings reported for FCC crystals~\cite{jasperson_cross-scale_2025}.
However, because the present study focuses on a BCC system and is based on a limited number of samples, these correlations should be considered suggestive rather than conclusive.

Although the strength prediction uncertainties vary substantially across subcases---with two subcases exhibiting spuriously large inflated uncertainties---they nonetheless overlap in a relatively narrow range.
All inflated uncertainties include the interval of approximately 2.19--2.38~GPa.
While direct DFT validation of the strength is infeasible, this overlapping interval provides a reasonable estimate of the true strength.
Encouragingly, this range also aligns with the values obtained from the proxy GT potentials.
We therefore propose this overlapping region as our best estimate of the plastic strength predicted by the DFT-fitted models.


\section{Discussion}
\label{sec:discussion}

\subsection{Model error correction}
\label{sec:model-error}

The results presented in Sec.~\ref{sec:results} highlight the effectiveness of the ALIM method in selecting small yet highly informative training sets that capture the parameter sensitivities necessary to predict indicator properties with specified precision.
However, despite these strengths, we observe that the resulting prediction uncertainties often underestimate the actual errors.
This limitation stems from the fact that ALIM guarantees precision---meaning it constrains parameter uncertainty relative to the chosen model---but cannot ensure accuracy if the model form itself cannot adequately represent the ground truth.
In other words, the method is dependent on the learning capacity of the chosen IP, and any inherent model error is not automatically included in the uncertainty estimates.
We expect the use of highly-flexible functional forms, such as machine-learned interatomic potentials, could avoid such issues, but at the expense of requiring a large number of parameters to be handled.

Our study demonstrates that in realistic IP development, model error can be substantial and, if left uncorrected, can lead to overconfident predictions of material properties.
A practical remedy is to apply a \emph{post hoc} uncertainty inflation correction, in which the propagated uncertainty estimates are rescaled by a model-error scale inferred from the fitting residuals or training loss~\cite{frederiksen_bayesian_2004,pernot_parameter_2017,kurniawan_bayesian_2022}.
This correction accounts for the fact that the residual error may reflect not only uncertainty in the fitted parameters, but also limitations of the model form itself.
Applying this correction generally improves the reliability of uncertainty estimates, making them better aligned with the observed errors.
Nevertheless, in cases with particularly large model errors, the corrected uncertainties can become so large that the predictions and uncertainties themselves become impractical.
This highlights that while such corrections can mitigate overconfidence, they do not overcome the limitations of an insufficiently flexible model. \textcolor{black}{The uncertainty inflation factor is estimated from the aggregate fitting residuals and is therefore a single global correction. As a result, it can be overly conservative for some quantities of interest and less accurate for others, especially when model error varies across different configurations or observables.}

Large-scale property predictions, such as plastic strength, present additional challenges due to the high cost or infeasibility of obtaining ground truth data.
In these applications, uncertainty estimates often serve as the only available proxy for model reliability.
Applying model error corrections is particularly important in this context to avoid underestimating prediction error and creating a false sense of confidence.
However, when the corrections are large, the resulting uncertainties can overwhelm the predictions, rendering them effectively unusable.
This once again illustrates how limitations in model flexibility directly impact the interpretability and practical utility of uncertainty estimates for large-scale properties.

\subsection{Sufficiency of indicator properties}
\label{sec:intermediate-qois}

While the central objective of this study is to develop IPs suitable for simulating plastic strength, we adopt a practical strategy that targets indicator properties that are expected to correlate with plastic strength to overcome the computational cost of plastic strength simulations.
Our results show predictions and associated uncertainties for these indicator properties and for plastic strength itself.
However, an important question remains: have we chosen indicator properties that contain sufficient information to effectively constrain plastic strength predictions?

To address this question, we perform an additional IM analysis after the ALIM iterations, this time comparing the FIMs of the five indicator properties and the plastic strength property.
Unlike the IM procedure used during potential development---where the candidate data are matched to the indicator properties---here we treat the indicator properties as the ``candidates'' and the plastic strength as the ``target QoI,'' performing the matching once at the end rather than iteratively.

As an illustration, we perform this analysis to the MD--EAM-proxy case.
However, the weight optimization problem is infeasible, meaning that no set of weights satisfies all constraints in Eq.~(\ref{eq:information-matching}).
This outcome indicates that the indicator properties are insufficient surrogates for the target QoI; consequently, matching the FIM to these indicator properties does not, in principle, guarantee sufficient information for predicting plastic strength.
Nevertheless, the training data selected using these indicator properties contain more information than is strictly required by the indicator properties themselves.
In particular, the FIM of the selected data also satisfies the information requirement for plastic strength, yielding finite and relatively small uncertainty in the predicted strength.
Thus, although the indicator properties are not strictly sufficient representations of the target QoI, information sufficiency for the target may still be achieved in practice.

To better understand this apparent success, we analyze how much of the plastic strength information is captured by the indicator properties.
Specifically, we quantify their shared information content by computing the sum of squared dot products between the eigenvectors of the plastic strength FIM and those spanning the range of the indicator properties’ FIM from the final ALIM iteration.
This metric provides a measure of how much of the plastic strength FIM’s eigenstructure---the type of required information---lies within the eigensubspace of the indicator properties’ FIM, i.e., the parameter directions that the indicator properties can, in principle, constrain.

Across all models, this analysis reveals that the indicator properties capture a substantial portion of the information required to constrain the target plastic strength.
In each case, over $86\%$ of the relevant eigenstructure is recovered, indicating that the selected indicator properties provide strong, albeit incomplete, coverage of the necessary parameter sensitivities.
For the MD--EAM-proxy case specifically, the indicator properties account for approximately $99\%$ of the required information.
The remaining information resides in the nullspace of the indicator properties’ FIM, with associated eigenvectors pointing to limited sensitivity to certain parameter combinations---particularly those involving $B$, $\eta$, and $F_e$.
These combinations reflect weak constraints on long-range attractive interactions and the embedding energy parameters.
Nevertheless, because the missing information is relatively small, it is plausible that the training data incidentally recovers this gap.

However, this level of success is not guaranteed, as it relies on incidental recovery of information missing from the chosen indicator properties.
To reduce this reliance, one can perform the IM analysis discussed above---matching indicator properties and target QoIs---prior to running the ALIM loop, ensuring that the indicator properties are explicitly chosen to cover the relevant parameter directions.
Equally important, expanding the set of indicator properties can further improve robustness; for example, vacancy formation and migration energies are promising candidates due to their physical relevance and demonstrated correlation with plastic strength in FCC metals \cite{jasperson_cross-scale_2025}.
By combining these strategies---early IM analysis and thoughtful expansion of indicator properties---the likelihood of satisfying the IM condition increases and lead to more reliable predictions.


\section{Conclusion}
\label{sec:conclusion}

In this study, we have demonstrated that the ALIM approach provides a powerful framework for developing bespoke IPs for specific target properties with specified uncertainty.
The central idea of ALIM is to select training data that provide the specific parameter space information required to achieve prescribed uncertainty targets for the QoIs, rather than simply maximizing global information content or reducing parameter uncertainty indiscriminately.
By systematically selecting a small yet informative subset of training data, the method enables precise constraints on model parameters relevant to the target QoIs.
However, to ensure reliable uncertainty quantification, it is critical to account for model error.
Without this correction, even optimally chosen training data cannot guarantee trustworthy uncertainty estimates when the model's representational limitations are significant.
We show that applying a \emph{post hoc} correction based on uncertainty inflation, which rescales the propagated uncertainties using a model-error scale inferred from the fitting residuals, can mitigate overconfident predictions.
However, in cases with substantial model error, this correction may lead to large uncertainties---highlighting fundamental limits to model reliability.

To address the challenge of high computational cost associated with large-scale target properties, we also explored an alternative strategy that targets computationally cheaper indicator properties that are expected to correlate with the large-scale target QoIs.
While our results suggest this is a promising approach, its success depends on whether the chosen indicator properties impose sufficient information constraints on the parameters relevant for predicting the large-scale property.
Even when they are formally insufficient, the selected training data can sometimes incidentally supply the missing information, although this outcome is not guaranteed in general.
Rather than relying on this luck, we recommend performing an initial analysis, such as IM-based sufficiency check as discussed in Sec.~\ref{sec:intermediate-qois}, prior to running ALIM iterations to ensure that the indicator properties can impose sufficient parameter constraint.

Finally, our results highlight that the accuracy and reliability of IPs are ultimately constrained by model expressiveness.
The limited flexibility of the EAM functional form emerges as a key source of both prediction error and uncertainty---affecting not only the matched QoIs but also complex target properties such as plastic strength.
We anticipate that both prediction errors and uncertainties can be reduced by adopting more flexible functional forms that better approximate the ground truth.
In particular, potentials based on complete and systematically extensible structural descriptors---such as Atomic Cluster Expansion (ACE) \cite{drautz_atomic_2019} or Moment Tensor Potentials (MTP) \cite{shapeev_moment_2016}---offer a principled path toward improved accuracy and uncertainty in the prediction of complex material properties.
Embedding such models within the ALIM framework would enable tighter parameter constraints, better predictive performance, and more rigorous uncertainty control via the FIM. \textcolor{black}{Although the present study uses a simple EAM form with only seven variable parameters to keep the information-matching problem tractable, ALIM framework could be extended to more complex machine-learned interatomic potentials. In such cases, matrix-free or approximate low-rank methods can reduce the cost of calculations involved in information matching.}


\backmatter
\bmhead{Acknowledgements}
Calculations were performed on the Borax and Lassen clusters at the Livermore Computing facility at LLNL.

\section*{Declarations}
\subsection*{Funding}
This work was performed under the auspices of the U.S. Department of Energy by Lawrence Livermore National Laboratory (LLNL) under Contract DE-AC52-07NA27344, and was funded by the Laboratory Directed Research and Development Program at LLNL under project tracking code 23-ERD-006.

\subsection*{Conflict of interest}
The authors declare no conflict of interests.

\subsection*{Data and code availability}
The data that support the findings of this study are openly available in Zenodo at \href{https://zenodo.org/records/18012696}{https://zenodo.org/records/18012696}.
The corresponding code is available at the following GitHub repository: \href{https://github.com/yonatank93/information-matching_eam-strength}{https://github.com/yonatank93/information-matching\_eam-strength}.

\subsection*{Author contributions}
Y.K. conceived and carried out the research, performed the investigation and analysis, and wrote the original draft.
Y.K. and M.K.T. developed the active learning method.
L.D.W. and D.S.-K. performed atomic environment extraction and provided atomic configurations.
L.D.W. also provided SNAP potential code and contributed to the investigation and analysis.
A.S. provided additional atomic configurations and the corresponding DFT ground-truth data.
I.N. and E.B.T. provided support with the OpenKIM infrastructure.
V.L. managed the project, acquired funding, and provided technical input.
V.V.B. provided the EAM potential code, carried out the plastic strength simulations, supervised the project, and contributed to writing the original draft.
All authors reviewed and edited the manuscript and approved the final version.

\bibliography{refs,refs_zotero}

\end{document}